# Multipartite Entanglement and Hypergraph states of three qubits

Ri Qu, Zong-shang Li, Juan Wang, Yan-ru Bao

School of Computer Science and Technology, Tianjin University, Tianjin, China, 300072

Tianjin Key Laboratory of Cognitive Computing and Application, Tianjin, China, 300072

Several entanglement measures are used to define equivalence classes in the set of hypergraph states of three qubits. Our classifications reveal that (i) under local unitary transformations, hypergraph states of three qubits are split into six classes and only one class of them is not equivalent to any graph state; (ii) under stochastic local operations with classical communication, for the single copy case hypergraph states of three qubits, partitioned into five classes which can not be converted into a W state, are equivalent to graph states; and (iii) when bipartite entanglement in three qubits considered, hypergraph states of three qubits are split into five classes and only one class of them has the same entangled graph as the W state.

1. Introduction

Any *graph state* [1-4] can be constructed on the basis of a (simple and undirected) graph. Although graph states can describe a large family of entangled states including *cluster states* [5], *GHZ states*, *stabilizer states* [6], etc., it is clear that they cannot represent all entangled states. To go beyond graph states and still keep the appealing connection to graphs, Ref. [7] introduces an axiomatic framework for mapping graphs to quantum states of a suitable physical system, and extends this framework to directed graphs and weighted graphs. Several classes of multipartite entangled states, such as *qudit graph states* [8], *Gaussian cluster states* [9], *projected entangled pair states* [10], and *quantum random networks* [11], emerge from the axiomatic framework. In [12], we generalize the above axiomatic framework to encoding hypergraphs into so-called quantum hypergraph states.

It has been known that hypergraph states include graph states [12]. One may ask whether hypergraph states are equivalent to graph states under local unitary transformations or stochastic local operations with classical communication (SLOCC), that is, whether hypergraph states can describe more quantum states than graph states under local unitary transformations or SLOCC. The main aim of this work is to answer the above question for the single copy case of three-qubit hypergraph states. For this, we will address the issue of quantifying and characterizing the entanglement of three-qubit hypergraph states by means of several bipartite and tripartite entanglement measures including *concurrence* [13], *entopic measure* [14], *3-tangle measure* [13] and *Schmidt measure* [15]. Several literatures have shown that there are several classifications of three-qubit pure states by means of the above measures. In [16], for the single case all three-qubit pure states are split into six classes, respectively called *A-B-C*, *A-BC*, *B-AC*, *C-AB*, GHZ-type and W-type, by SLOCC. Ref. [17] partition all pure states of three qubits into eight classes [18] by means of the so-called *entangled graph*. In this paper, we will define different equivalence classes in the set of three-qubit hypergraph states according to the approaches shown in [16] and [17]. Our classifications reveal that (i) a class of three-qubit hypergraph states is not equivalent to any graph state under local unitary transformations; (ii) any state in the above class is equivalent to a graph state (up to local unitaris equivalent to the GHZ state) under SLOCC, which implies hypergraph states can not describe the W-type class [16] including a *W state*; and (iii) when bipartite entanglement in three qubits is considered, any state in the above class is of W-like states [17] where each qubit pair is entangled like the W state, i.e., its entangled graph contains three edges.

This paper is organized as follows. In Sec. 2, we recall notations of hypergraphs, hypergraph states, etc. In Sec. 3, we quantify the entanglement of three-qubit hypergraph states by means of local entropic measures and prove the existence of six classes of hypergraph states of three qubits under local unitary transforms. In Sec. 4, we quantify the entanglement of genuine tripartite entangled hypergraph states of three qubits by means of 3-tangle measure and spilt three-qubit hypergraph states into five classes under SLOCC. We also indicate that any hypergraph state can not be converted into the W state by SLOCC. In Sec. 5, we evaluate the entanglement of hypergraph states of three qubits by means of the Schmidt measure. In Sec. 6, we discuss bipartite entanglement of hypergraph states of three qubits by the concurrence and draw the corresponding entangled graphs. Section 7 contains our conclusions.

2. Preliminaries

Formally, a *hypergraph* is a pair $(V, E)$, where $V$ is the set of *vertices*, $E \subseteq \wp(V)$ is the set of *hyperedges* and $\wp(V)$ denotes the power set of the set $V$. Let $V \equiv \{A, B, C\}$ since we only consider three-vertex hypergraphs in this paper. Moreover, for a set of the hyperedges $F \subseteq \wp(V)$, adding all hyperedges of $F$ to a hypergraph $g = (V, E)$ will obtain a new hypergraph $g + F \equiv (V, E \Delta F)$ where $E \Delta F$ denotes the symmetric difference of $E$ and $F$, that is, $E \Delta F = E \bigcup F - E \bigcap F$.

Let $I$ be the 2 by 2 identity matrix and $Z_k$ be the $2^k$ by $2^k$ diagonal matrix which satisfies

$$(Z_k)_{jj} = \begin{cases} -1 & j = 2^k \\ 1 & others \end{cases} \quad (1)$$

where $k \in \{0, 1, 2, 3\}$. Suppose $e \subseteq V$. Then the three-qubit *hypergraph gate* $Z_e$ is defined as $Z_{|e|} \otimes I^{\otimes 3-|e|}$ which means that $Z_{|e|}$ acts on the qubits in $e$ while the identity $I$ acts on the rest respectively.

A three-qubit *hypergraph state* $|g\rangle$ can be constructed by $g = (V, E)$ as follows. Each vertex labels a qubit (associated with a Hilbert space $\mathbb{C}^2$) initialized in $|\phi\rangle = |+\rangle \equiv \frac{1}{\sqrt{2}}(|0\rangle + |1\rangle)$. The state $|g\rangle$ is obtained from the initial state $|+\rangle^{\otimes 3}$ by applying the hyperedge operator $Z_e$ for each hyperedge $e \in E$, that is,

$$|g\rangle = \prod_{e \in E} Z_e |+\rangle^{\otimes 3}. \tag{2}$$

Thus hypergraph states of three qubits are corresponding to $\left(\mathbb{C}^2, |+\rangle, \{Z_k \mid 0 \leq k \leq 3\}\right)$ by the axiomatic approach while graph states are related with $\left(\mathbb{C}^2, |+\rangle, Z_2\right)$ [7, 12].

It is known that real equally weighted states [19] are equivalent to hypergraph states [12]. In fact, define a mapping $c$ on $\wp(V)$ as

$$\forall e \subseteq V, c(e) = \begin{cases} 1 & e = \Phi \\ \prod_{k \in e} x_k & e \neq \Phi \end{cases}. \tag{2}$$

Then we can construct a *1-1* mapping $u$ between hypergraphs and Boolean functions which satisfies $\forall g = (V, E)$,

$$u(g) = \bigoplus_{e \in E} c(e). \tag{3}$$

where $\oplus$ denotes the addition operator over $\mathbb{Z}_2$. Thus we have

$$|g\rangle = \prod_{e \in E} Z_e |+\rangle^{\otimes 3} = \frac{1}{\sqrt{2^3}} \sum_{x=0}^{2^3-1} (-1)^{\bigoplus_{e \in E} c(e)} |x\rangle \equiv |\psi_{u(g)}\rangle \tag{4}$$

where $|\psi_{u(g)}\rangle$ is just the real equally weighted state associate with the Boolean function $u(g)$.

For the single case it is known two pure states can be obtained with certainty from by means of LOCC if and only if they are related by local unitaries [14]. Moreover, they can be converted by means of SLOCC if and only if they are associated with an invertible local operator [16]. Let *g* and *g'* be two hypergraphs. We say that they are *LU-equivalent*, if there exists a local unitary *U* such that

$$|g\rangle = U|g'\rangle, \tag{5}$$

i.e., $|g\rangle$ and $|g'\rangle$ are equivalent under local unitary operations. If there exists an invertible local operator *O* such that

$$|g\rangle = O|g'\rangle, \tag{6}$$

that is, $|g\rangle$ and $|g'\rangle$ are equivalent under SLOCC, then we say that *g* and *g'* are *SLOCC-equivalent*.

3. Entropic measure and LU-equivalent classes

Given three qubits *A*, *B* and *C*, we can regard the three-qubit system as a bipartite system. For instance, *A* is one part of the system and the remaining tow qubits *B* and *C* is the other.

Correspondingly, a pure state $|\phi\rangle$ of three qubits can be viewed as a bipartite state $|\phi_{A(BC)}\rangle$. The local entropic measure $E_2^A(|\phi\rangle)$ is given by the smallest eigenvalue of the reduced density matrix $\rho_A \equiv Tr_{BC}(|\phi\rangle\langle\phi|)$. Similarly, we also can define $E_2^B(|\phi\rangle)$ and $E_2^C(|\phi\rangle)$. It is known the local entropic measures are an entanglement monotone for a single copy case and invariant under local unitary operations.

*Proposition 1.* All three-vertex hypergraphs are split into six LU-equivalence classes as follows.

$$G_0 = \{(V,E) \mid E \in \wp(\{\{\Phi\},\{A\},\{B\},\{C\}\})\}, \quad G_1 = \{g + \{\{B,C\}\} \mid g \in G_0\},$$

$$G_2 = \{g + \{\{A,C\}\} \mid g \in G_0\}, \quad G_3 = \{g + \{\{A,B\}\} \mid g \in G_0\},$$

$$G_4 = \{g + E \mid E \subseteq \{\{A,B\},\{A,C\},\{B,C\}\} \wedge |E| \geq 2 \wedge g \in G_0\},$$

$$G_5 = \{(V,E) \mid V \in E\}. \tag{7}$$

*Proof.* We first prove that for $0 \leq k \leq 5$ any two hypergraphs in $G_k$ are LU-equivalent. Let $g_0 \equiv (V,\Phi)$, $g_1 \equiv (V,\{\{B,C\}\})$, $g_2 \equiv (V,\{\{A,C\}\})$ and $g_3 \equiv (V,\{\{A,B\}\})$. It is clear for $0 \leq k \leq 3$ that the hypergraph $g_k$ is LU-equivalent to any hypergraph in $G_k$. Let $g_4 \equiv (V,\{\{A,B\},\{B,C\},\{A,C\}\})$. It is known that $g_4$ and $(V,\{\{A,B\},\{A,C\}\})$ are LU-equivalent [1]. Thus the hypergraph $g_4$ is LU-equivalent to any hypergraph in $G_4$. Let $g_5 \equiv (V,\{V\})$. According to proposition 2 in [12], the hypergraph $g_5$ is also LU-equivalent to any hypergraph in $G_5$.

Now we prove that any two hypergraphs in $\{g_k \mid 0 \leq k \leq 5\}$ are not LU-equivalent by means of local entropic measures. For any $g = (V,E)$, it is known that $|g\rangle = |\psi_{u(g)}\rangle$ by (4). Thus the reduced density matrix $\rho_A(g)$ of $|g\rangle$ is obtained by

$$\rho_A(g) = Tr_{BC}(|g\rangle\langle g|) = [a_{ij}^{(A)}]_{2\times 2} \tag{8}$$

where $a_{ij}^{(A)} = \frac{1}{8}\sum_{x_B,x_C=0}^{1}(-1)^{u(g)(i,x_B,x_C)\oplus u(g)(j,x_B,x_C)}$. By simple computation, we can have

$$\rho_A(g) = \begin{bmatrix} \frac{1}{2} & a \\ a & \frac{1}{2} \end{bmatrix} \quad (9)$$

where $a \equiv a_{01}^{(A)} = \frac{1}{8} \sum_{x_B, x_C = 0}^{1} (-1)^{u(g)(0, x_B, x_C) \oplus u(g)(1, x_B, x_C)}$. According to (3), we can get

$$u(g_0)(x_A, x_B, x_C) = 0, \quad u(g_1)(x_A, x_B, x_C) = x_B x_C, \quad u(g_2)(x_A, x_B, x_C) = x_A x_C,$$

$$u(g_3)(x_A, x_B, x_C) = x_A x_B, \quad u(g_4)(x_A, x_B, x_C) = x_A x_B \oplus x_A x_C \oplus x_B x_C,$$

$$u(g_5)(x_A, x_B, x_C) = x_A x_B x_C. \quad (10)$$

Thus it is easy to obtain the values of the local entropic measures of $E_2^A$, $E_2^B$ and $E_2^C$ for all states in $\{|g_k\rangle \mid 0 \leq k \leq 5\}$. These values are shown in Table 1. Clearly, any two hypergraphs in $\{g_k \mid 0 \leq k \leq 5\}$ are not LU-equivalent since the local entropic measures are invariant under local unitary operations. □

Table 1. Values of several entanglement measures associated with different hypergraph subsets.

|       | $E_2^A$ | $E_2^B$ | $E_2^C$ | $\tau$ | $C_{AB}$ | $C_{AC}$ | $C_{BC}$ |
|-------|---------|---------|---------|--------|----------|----------|----------|
| $G_0$ | 0 | 0 | 0 | 0 | 0 | 0 | 0 |
| $G_1$ | 0 | $\frac{1}{2}$ | $\frac{1}{2}$ | 0 | 0 | 0 | 1 |
| $G_2$ | $\frac{1}{2}$ | 0 | $\frac{1}{2}$ | 0 | 0 | 1 | 0 |
| $G_3$ | $\frac{1}{2}$ | $\frac{1}{2}$ | 0 | 0 | 1 | 0 | 0 |
| $G_4$ | $\frac{1}{2}$ | $\frac{1}{2}$ | $\frac{1}{2}$ | 1 | 0 | 0 | 0 |
| $G_5$ | $\frac{1}{4}$ | $\frac{1}{4}$ | $\frac{1}{4}$ | $\frac{1}{4}$ | $\frac{1}{2}$ | $\frac{1}{2}$ | $\frac{1}{2}$ |

Note that the state $|g_4\rangle$ is up to local unitaries equivalent to the GHZ state. Clearly, any three-vertex (simple and undirected) graph belongs to one of $G_0$, $G_1$, $G_2$, $G_3$ and $G_4$ while it is not of $G_5$. From the above proposition, it is obtained that any state associated with $G_5$ is not equivalent to graph states under local unitary transformations.

4. 3-tangle measure and SLOCC-equivalent classes

Let a three-qubit pure state $|\phi\rangle = \sum_{i,j,k=0}^{1} a_{ijk} |ijk\rangle$. The 3-tangle measure $\tau(|\phi\rangle)$ is given by

$$\tau(|\phi\rangle) = 2\left|\sum a_{ijk} a_{i'j'm} a_{npk'} a_{n'p'm'} \varepsilon_{ii'} \varepsilon_{jj'} \varepsilon_{kk'} \varepsilon_{mm'} \varepsilon_{nn'} \varepsilon_{pp'}\right| \tag{11}$$

where $\varepsilon_{01} = -\varepsilon_{10} = 1$ and $\varepsilon_{00} = \varepsilon_{11} = 0$ [13]. It is known the measure is an entanglement monotone and invariant under local unitary transformations. Moreover, 3-tangle measure is also invariant under permutations of the qubits.

*Proposition 2.* All three-qubit hypergraph states are partitioned into five classes under SLOCC, that is, in the set of three-vertex hypergraphs there exist five SLOCC-equivalence classes $G_0$, $G_1$, $G_2$, $G_3$, and $G_4 \cup G_5$ which are respectively associated with the classes *A-B-C*, *A-BC*, *B-AC*, *C-AB* and GHZ-type defined in [16].

*Proof.* By (10) and (11), it is easy to obtain $\tau(|g_k\rangle) = 0 \, (0 \le k \le 3)$, $\tau(|g_4\rangle) = 1$ and $\tau(|g_5\rangle) = \frac{1}{4}$. According to the method shown in [16], we can use the values of the local entropic measures and 3-tangle measure to obtain that three-vertex hypergraphs are split into five SLOCC-equivalence classes: $G_0$, $G_1$, $G_2$, $G_3$, and $G_4 \cup G_5$ which are corresponding to the classes *A-B-C*, *A-BC*, *B-AC*, *C-AB* and GHZ-type. □

Clearly, three-vertex hypergraphs are SLOCC-equivalent to graphs of three vertices, this is, three-qubit hypergraph states are equivalent to graph states of three qubits under SLOCC. In particular, any hypergraph state associated with $G_4 \cup G_5$ can be converted into the GHZ state by means of SLOCC. According to the above proposition, any hypergraph state of three qubits is not of the W-type class defined in [16]. Thus we can obtain the following corollary, which implies that hypergraph states do not represent all states of three qubits.

*Corollary 3.* Any three-qubit hypergraph state can not be converted into the W state by SLOCC.

5. Schmidt measure

Any pure state $|\phi\rangle$ of three qubits can be represented as

$$|\phi\rangle = \sum_{i=1}^{R} a_i |\phi_i^{(A)}\rangle \otimes |\phi_i^{(B)}\rangle \otimes |\phi_i^{(C)}\rangle \tag{12}$$

where $a_i \in \mathbb{C}(i=1,2,...,R)$, and $|\phi_i^{(A)}\rangle$, $|\phi_i^{(B)}\rangle$ and $|\phi_i^{(C)}\rangle$ are one-qubit pure states. The Schmidt measure of $|\phi\rangle$ is defined as

$$E_S(|\phi\rangle) = \log_2(r) \tag{13}$$

where $r$ is the minimal number $R$ of terms in the sum of (12) over all linear decompositions into product states. It is known the measure is an entanglement monotone and invariant under local unitary transformations.

*Proposition 4.* The Schmidt measure of any three-quibt hypergraph state is either *0* or *1*.

*Proof.* Ref. [16] has shown that the Schmidt measure $E_S$ of any state in the class *A-B-C* is *0* while $E_S$ of any state in the classes *A-BC*, *B-AC*, *C-AB* and GHZ-type is *1*. However, the Schmidt measure of any state in the W-type class is equal to $\log_2(3)$. According to corollary *3*, the Schmidt measure of any three-quibt hypergraph state is equal to one of *0* and *1*. □

6. Concurrence and entangled graphs

Concurrence is a famous bipartite entanglement measure. Let $|\phi\rangle$ be a pure state of three qubits *A*, *B* and *C*. The reduced density matrix $\rho_{AB}$ of $|\phi\rangle$ is defined as $\rho_{AB} \equiv Tr_C(|\phi\rangle\langle\phi|)$. One can evaluate the so-called spin-flipped operator defined as

$$\tilde{\rho}_{AB} = (\sigma_y \otimes \sigma_y)\rho_{AB}^*(\sigma_y \otimes \sigma_y) \tag{14}$$

where $\sigma_y$ is the Pauli matrix and a star denotes a complex conjugation. Let $\lambda_1$, $\lambda_2$, $\lambda_3$ and $\lambda_4$ be eigenvalues of the matrix $\rho_{AB}\tilde{\rho}_{AB}$ in decreasing order. The concurrence $C_{AB}$ between two qubits *A* and *B* is defined as

$$C_{AB} \equiv \max\{0, \sqrt{\lambda_1} - \sqrt{\lambda_2} - \sqrt{\lambda_3} - \sqrt{\lambda_4}\} \tag{15}$$

It is known that $\rho_{AB}$ is separable or disentangled if and only if $C_{AB} = 0$. Moreover, the concurrence $C_{A(BC)}$ between one qubit *A* and the other two qubits is equal to $2\sqrt{\det \rho_A}$ where $\rho_A = Tr_{BC}(|\phi\rangle\langle\phi|)$. Thus we can obtain

$$C_{A(BC)} = 2\sqrt{E_2^A \cdot (1 - E_2^A)}. \tag{16}$$

Similarly, we can also define $C_{AC}$, $C_{BC}$, $C_{B(AC)}$ and $C_{C(AB)}$.

From (16), we can evaluate the values of $C_{A(BC)}$, $C_{B(AC)}$ and $C_{C(AB)}$ for three-qubit hypergraph states by using local entropic measures obtained in Sec. 3. Since $\tau \equiv \tau_{ABC} = C^2_{A(BC)} - C^2_{AB} - C^2_{AC}$ (which is the original definition of 3-tangle measure [13]) and $\tau$ is variant under permutations of the qubits, it is easy to obtain the values of $C_{AB}$, $C_{AC}$ and $C_{BC}$ for three-qubit hypergraph states. These values are shown in Table 1.

Ref. [17] introduces a concept of an entangled graph such that each qubit of a multipartite system is associated with a vertex, while a bipartite entanglement between two specific qubits is represented by an edge between these vertices. For an *n*-qubit state, its entangled graph can visually show how a bipartite entanglement is "distributed" in *n* qubits. According to $C_{AB}$, $C_{AC}$ and $C_{BC}$, we can draw entangled graphs of three-qubit hypergraph states, which are shown in Figure 1. Thus all hypergraph states of three qubits are classified into five classes as follows. The entangled graph associated with $G_0 \cup G_4$ has no edge while one related with $G_1 \cup G_2 \cup G_3$ has only one edge. Moreover, the entangled graph corresponding to $G_5$ is a complete graph with three vertices, which means hypergraph states associated with $G_5$ are of the W-like states. However, hypergraph states of three qubits do not conclude the states whose entangled graphs have just two edges. This implies that hypergraph states do not represent all states of three qubits again.

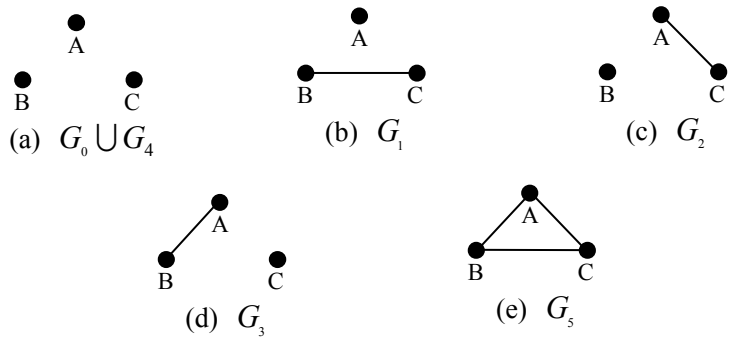

Figure 1. Entangled graphs associated with three-vertex hypergraphs.

7. Conclusions

This work uses several bipartite and tripartite entanglement measures to quantify and characterize the entanglement of hypergraph states of three qubits, as shown in Table 1. According to the values of these measures, we define the equivalence classes of hypergraph states and prove that the states associated with $G_5$ are not equivalent to any graph state under local unitary transformations. However, hypergraph states of three qubits are equivalent to graph states under SLOCC. And any hypergraph state of three qubits can not be converted into the W state by SLOCC. Thus hypergraph states can not represent all pure states of three qubits. Moreover, when bipartite entanglement in three qubits considered, the states corresponding to $G_5$ are related with an entangled graph which contains three edges while entangled graphs of graph states include at most one edge. Although any hypergraph state associated with $G_5$ is not (up to local unitaries or SLOCC) equivalent to the W state, its every qubit pair is entangled like W state. This property of bipartite entanglement of the W state has been used in some quantum information processing tasks. Thus it is helpful for these tasks that the W state is replaced to the state $|g_5\rangle$ which might be prepared more easily than the W state in some cases.

## ACKNOWLEDGMENTS

This work was financially supported by the National Natural Science Foundation of China under Grants No. 61170178 and No. 61272254.**References**
[1] M. Hein, J. Eisert, and H. J. Briegel, Phys. Rev. A 69, 062311 (2004).
[2] H. Aschauer, W. Dur, and H. J. Briegel, Phys. Rev. A 71, 012319 (2005).
[3] R. Raussendorf, D. E. Browne, and H. J. Briegel, Phys. Rev. A 68, 022312 (2003).
[4] R. Raussendorf and H. J. Briegel, Phys. Rev. Lett. 86, 5188 (2001).
[5] H. J. Briegel and R. Raussendorf, Phys. Rev. Lett. 86, 910 (2001).
[6] D. Gottesman, Phys. Rev. A 54, 1862 (1996); D. Gottesman, Phys. Rev. A 57, 127 (1998).
[7] R. Ionicioiu and T. P. Spiller, Phys. Rev. A 85, 062313 (2012).
[8] S. Y. Looi *et al.*, Phys. Rev. A 78, 042303 (2008).
[9] N. C. Menicucci, S. T. Flammia, and P. van Loock, Phys. Rev. A 83, 042335 (2011).
[10] F. Verstraete *et al.*, Phys. Rev. Lett. 96, 220601 (2006); D. Pérez-García *et al.*, Quantum Inf. Comput. 8, 650 (2008); N. Schuch *et al.*, Phys. Rev. Lett. 98, 140506 (2007).
[11] S. Perseguers *et al.*, Nat. Phys. 6, 539 (2010).
[12] Ri Qu *et al.*, e-print arXiv: 1211.3911 [quant-ph].
[13] V. Cffman, J. Kundu, and W. K. Wootters, Phys. Rev. A 61, 052306 (2000).
[14] G. Vidal, J. Mod. Opt. 47, 355 (2000).
[15] J. Eisert and H. J. Briegel, Phys. Rev. A 64, 022306 (2001).
[16] W. Dür, G. Vidal, and J. I. Cirac, Phys. Rev. A 62, 062314 (2000).
[17] M. Plesch and V. Bužek, Phys. Rev. A 67, 012322 (2003).
[18] In fact, Ref. [17] splits all pure states of three qubits into four classes since the entangled graph isomorphism is considered.
[19] D. Bruß and C. Macchiavello, Phys. Rev. A 83, 052313 (2011).that the states associated with $G_5$ are not equivalent to any graph state under local unitary transformations. However, hypergraph states of three qubits are equivalent to graph states under SLOCC. And any hypergraph state of three qubits can not be converted into the W state by SLOCC. Thus hypergraph states can not represent all pure states of three qubits. Moreover, when bipartite entanglement in three qubits considered, the states corresponding to $G_5$ are related with an entangled graph which contains three edges while entangled graphs of graph states include at most one edge. Although any hypergraph state associated with $G_5$ is not (up to local unitaries or SLOCC) equivalent to the W state, its every qubit pair is entangled like W state. This property of bipartite entanglement of the W state has been used in some quantum information processing tasks. Thus it is helpful for these tasks that the W state is replaced to the state $|g_5\rangle$ which might be prepared more easily than the W state in some cases.

## ACKNOWLEDGMENTS

This work was financially supported by the National Natural Science Foundation of China under Grants No. 61170178 and No. 61272254.

**References**
[1] M. Hein, J. Eisert, and H. J. Briegel, Phys. Rev. A 69, 062311 (2004).
[2] H. Aschauer, W. Dur, and H. J. Briegel, Phys. Rev. A 71, 012319 (2005).
[3] R. Raussendorf, D. E. Browne, and H. J. Briegel, Phys. Rev. A 68, 022312 (2003).
[4] R. Raussendorf and H. J. Briegel, Phys. Rev. Lett. 86, 5188 (2001).
[5] H. J. Briegel and R. Raussendorf, Phys. Rev. Lett. 86, 910 (2001).
[6] D. Gottesman, Phys. Rev. A 54, 1862 (1996); D. Gottesman, Phys. Rev. A 57, 127 (1998).
[7] R. Ionicioiu and T. P. Spiller, Phys. Rev. A 85, 062313 (2012).
[8] S. Y. Looi *et al.*, Phys. Rev. A 78, 042303 (2008).
[9] N. C. Menicucci, S. T. Flammia, and P. van Loock, Phys. Rev. A 83, 042335 (2011).
[10] F. Verstraete *et al.*, Phys. Rev. Lett. 96, 220601 (2006); D. Pérez-García *et al.*, Quantum Inf. Comput. 8, 650 (2008); N. Schuch *et al.*, Phys. Rev. Lett. 98, 140506 (2007).
[11] S. Perseguers *et al.*, Nat. Phys. 6, 539 (2010).
[12] Ri Qu *et al.*, e-print arXiv: 1211.3911 [quant-ph].
[13] V. Cffman, J. Kundu, and W. K. Wootters, Phys. Rev. A 61, 052306 (2000).
[14] G. Vidal, J. Mod. Opt. 47, 355 (2000).
[15] J. Eisert and H. J. Briegel, Phys. Rev. A 64, 022306 (2001).
[16] W. Dür, G. Vidal, and J. I. Cirac, Phys. Rev. A 62, 062314 (2000).
[17] M. Plesch and V. Bužek, Phys. Rev. A 67, 012322 (2003).
[18] In fact, Ref. [17] splits all pure states of three qubits into four classes since the entangled graph isomorphism is considered.
[19] D. Bruß and C. Macchiavello, Phys. Rev. A 83, 052313 (2011).